# Ring aggregation pattern of Human Travel Trips

Zi-Yang Wang[1,2], Wen-Yu Li[1], Peng Zhu[1], Yong Qin[2], Li-Min Jia[2]

[1]School of Traffic and Transportation, Beijing Jiaotong University, Beijing 100044, China
[2]State Key Laboratory of Railway Control and Safety, Beijing Jiaotong University, Beijing 100044, China

**Abstract** Although a lot of attentions have been paid to human mobility, the relationship between travel pattern with city structure is still unclear. Here we probe into this relationship by analyzing the metro passenger trip data. There are two unprecedented findings. One, from the average view a linear law exists between the individual's travel distance with his original distance to city center. The mechanism underlying is a travel pattern we called "ring aggregation", i.e. the daily movement of city passengers is just aggregating to a ring with roughly equal distance to city center. Interestingly, for the round trips the daily travel pattern can be regarded as a switching between the home ring at outer area with the office ring at the inner area. Second, this linear law and ring aggregation pattern seems to be an exclusive characteristic of the metro system. It can not be found in short distance transportation modes, such as bicycle and taxi, neither as multiple transportation modes. This means the ring aggregation pattern is a token of the relationship between travel pattern with city structure in the large scale space.

## 1. Introduction

Recently, Human mobility has attracted many interests from different fields for its potential relation with many applications, such as urban planning [1-4], traffic engineering [5-7], epidemics spreading [8–13] and location-based service [14-15]. The current studies mainly focused on the universal pattern of human mobility. The earlier research about dollar circulation revealed the human mobility is not simple or regular [16], but complex and heterogenic, which appears as the power law distribution of walk displacement. This work shed new light on the study of human mobility, and later, scientists found the same power law distribution from mobile phone data [17-19], GPS tail data [20-23], global ship cargo data [24-26] and so on. While recent research find that single mode of transportation will lead exponential distribution but not power law. For examples, [27] examined the GPS tail of Beijing taxi during three months and found the placement obeys exponential distribution, [28] found the 9000km-above-distance US flight tail among 11 days exhibits exponential distribution. In fact, as we examined in this work, the metro data of Beijing also shows exponential law.

So a natural question arises, which is the fundamental law of human mobility, the power law one or the exponential one? Recent research shows that power law may result from the mixture of different transportation modes [29]. And as an interesting discussion about the dairy travel data of volunteers, [30] believed that the mixed displacement is power law though each individual moves irregularly.

On one side we are benefited and inspired from the above discussions on the universal law of human mobility. While on the other side, it is a little tedious if all the discussions are concentrated on only one facet of human mobility.

In fact, many other factors may also affect human mobility, such as city structure. From the viewpoint of complex system, city is a kind of spatial network with the functional spots are the nodes and the traffic roads are the links [31]. As the theory of "structure decides its function" illuminates, city structure should also give a deep effect to the human travel pattern. And at the same time, the characteristics of city structure should also be mined from the travel tail of city passengers. While the relationship between human mobility with city structure is quite ignored from discussions, only some simple characteristics of city structure are related to the human mobility, such as city center [32].

In this work, we give a deep probe to the relationship between human mobility with city structure. We analyzed the daily ticket card data of Beijing metro system and found a linear relationship existing between the mean travel distance of individuals with the distance from their original spots to city center. We explained the mechanism behind this linear relationship by a pattern we named "ring aggregation", in other words the movement of passengers can be regarded as the action of aggregating to a ring, a zone with roughly equal distance to city center. So this linear law establishes a path from city structure to human travel pattern. As the knowledge of us it is the first time this law is found and it is reasonable believing this work will have a lot of potential applications in the future.

## 2. Phenomenon

First, as [32] narrated, we find the characteristics "city center" can indeed be reflected by passenger trip data. The method used in [32] is a clustering method. Here we use a more simple and direct way. We sort the total inflow of each station (the result is same for outflow). In figure 1 we show the first ten stations in (a), the first twenty stations in (b), etc. Then it is clear that big stations are more likely to locate near the center and small stations are more likely to locate at peripheral area. This shows that the center is indeed existing. In Supplementary Materials part 3 we offer a better proof for the existence of city center. We found the law is same with London tube system and Shenzhen (a developed city of China) metro system. According to the inflow and site of the station, we define the center by $P_c = [x_c, y_c] = \sum_{i=1}^{n} m_i [x_i, y_i] / \sum_{i=1}^{n} m_i$, where $n$ is the total number of stations, $[x_i, y_i]$ is the site of the station $i$ and $m_i$ is the total inflows of station $i$.

Figure 1. Top stations in the amount of inflow. From a to j the red squares are respectively the top ten, top 20, …, top 100 stations in the amount of inflow, the green squares are the stations of the whole metro system.

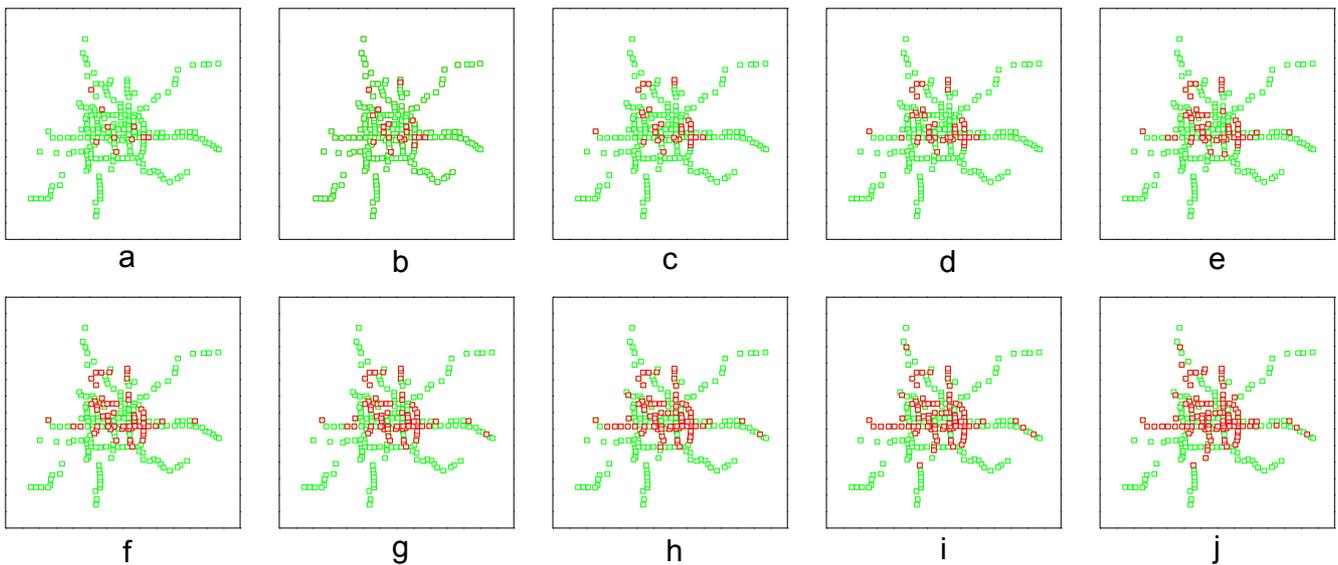

Second, more important, we found there is a quite well linear relationship between the mean travel distance of individuals with the distance from their original spots to city center, as figure 2 shows. This means the travel pattern is obviously related to city center. And it is worth speaking that to our knowledge until now no similar law is reported in the research of human mobility.

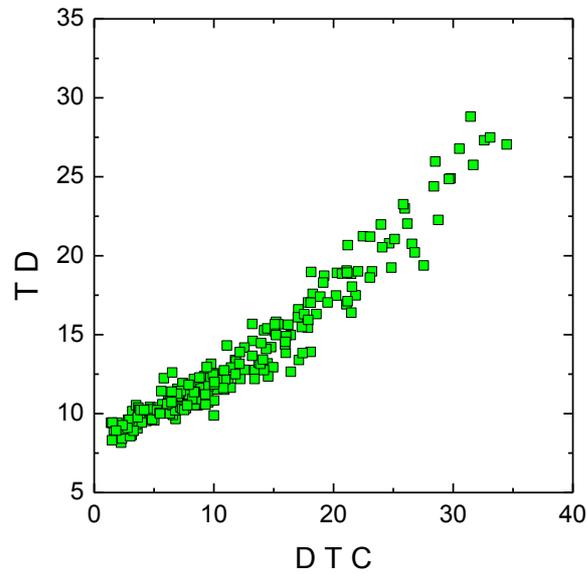

Figure 2. The linear law between the travel distance (TD) with the passengers' original distance to city center (DTC). Each square behaves as a station, the x coordinates is the distance to center of this station, and the y coordinates is the average travel distance of the inflow passengers of this station.

As we can see that this linear relationship is a law related to city structure, a question naturally arises. What is the mechanism behind this linear law? And which travel pattern and city structure it reflects?

## 3. Mechanics

The linear law in figure 2 means the farther distance to city center an individual starts, the farther he will travel, which implies city center has a dominant influence to human mobility. And it makes a natural guess that the travel pattern of city passengers is that all the passengers travel and aggregate to the city center. If this guess is tenable, the linear law will naturally hold. But the guess is easy to verify incorrect. It is impossible that city center is the only destination for all passengers. First, the city center can not congest unlimitedly. Second, there is a large part of round trip among daily passenger flow (the definition of round trip can see at Supplementary Materials part 2). If one part of round trip is aggregating to the city center, the other part with the same amount must be dispersing to the city periphery. It is impossible that all the round trips only move to the city center. Furthermore, for the dispersing part the linear law may be invalid. Then which pattern can make the whole of the round trip, both the aggregating part and the dispersing part to fit the linear law simultaneously?

To mend the above bug it is not difficult to update the travel pattern to the "proportional moving" pattern as shown in figure 3. Under this pattern the towards-center moving and the departing-from-center moving are all allowed. But for both directions, the individual travel distance should be proportional to the distance between his starting spot to city center. Then the linear law may be satisfied by both part of the round trip. For an example, as we can see in figure 3, A,A' and B,B' are two pair of round trips. Among the movement of both aggregating and dispersing, they fit the linear law.

But the "proportional moving" pattern is not belonging to the metro passengers either, because under that pattern a passenger with farther origin will also go to a farther destination, while we find this is not truth.

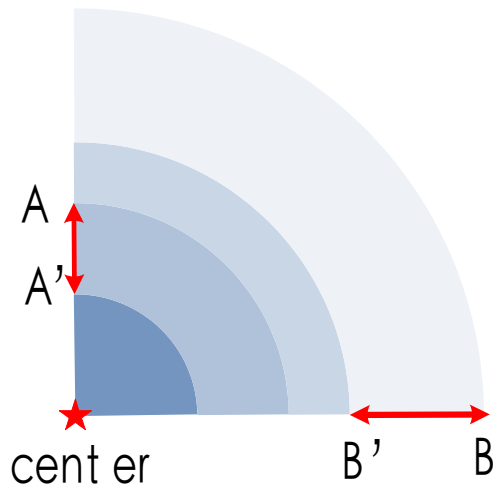

Figure 3. The "proportional moving" pattern. AA' and BB' are two pair of round trips. In the movement of both aggregating and dispersing, they fit the law that the farther to the city center an individual starts, the farther he will travel.

In order to disclose the travel pattern of metro trips more clearly, in the next step we analyzed where the passengers go by means of average, i.e. averaging the distance to city center of each station's outflow. Then for each station we get the distance to center of its average destination.

It is very clear that though before averaging the destinations of outflow are diversity and the stations are quite different in position, after averaging the destination of each station is astonishingly similar, all locate around the center with nearly equal distance. This implies by averaging view all the passengers aggregate to a ring with the almost equal distance to city center, show in the inset of figure 4. It is worthy to note that the distance from the ring to center coincide with the mean distance of all the outflows (the outflows of all the stations), they are all near the value 10.2647.

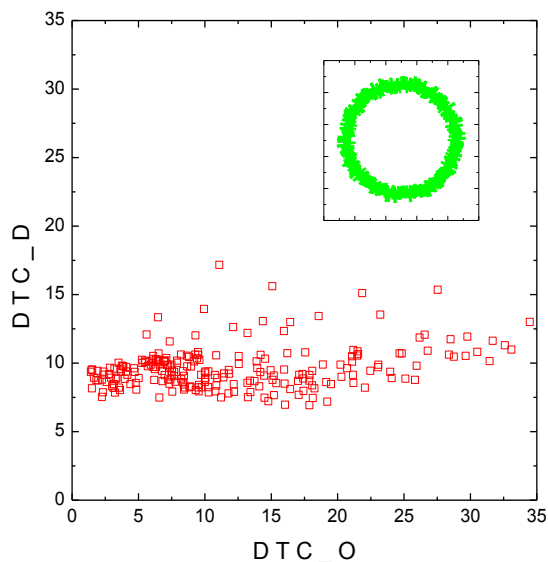

Figure 4. The average distance to center of the outflow of each station. Each square behaves a station, the DTC_O is the distance to center of this station and DTC_D is the average distance to center of the outflow of this station; The inset shows the ring aggregation pattern.

The ring aggregation pattern and the linear law in figure 2 can also be found in London tube system and Shenzhen metro system.(See Supplementary Materials part 4), which implies that the linear law is an universal law for metro.

Is the ring aggregation the travel pattern of metro passengers? Or in other word is there something else to prove the ring aggregation theory except the above discussion about Figure 4?

Under the travel pattern of ring aggregation, the travel distance of an individual will be proportional to his distance to the ring. Then if that pattern is tenable, two deduced results will appear. The first is the linear law appears better in the far away stations. The second is the closer the ring locates to the center, the better the linear law will be.

All the trip data (Beijing, Shenzhen, London) validate the first point, as we can see figure S2.a, figure S2.b in Supplementary Materials part 4. Especially we can see in figure S2.a in Supplementary Materials part 4, for London tube flow the line drops in a small scale near the center, which fits this deduction perfectly.

And the validation to the second point can see figure 5, it shows a powerful proof to the ring aggregation theory. We found after the daily round trips are divided into two parts, the departure flow and the return flow, both parts will aggregates to a ring and both parts obey the linear law, as figure 5 and figure 6 shows. The definition and the classify method of the two parts can see in the Supplementary Materials part 2.

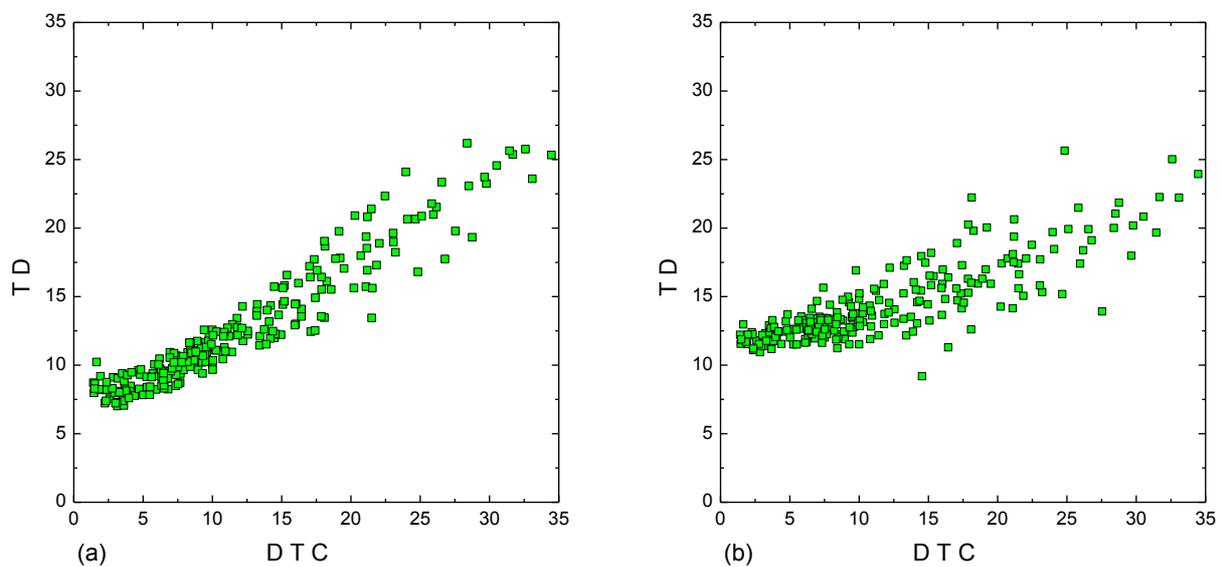

Figure 5. The linear law between the travel distance (TD) with the passengers' original distance to city center (DTC). (a) the departure flow, (b)the return flow. The symbols are same with figure 2.

Figure S1 in Supplementary Materials part 3 shows that the departure flow locates at periphery area while the return flow locates at the inner area. We find the same characteristic by means of ring. The ring for the departure flow also locates more periphery than the return flow does. Then we can see that the linear law for the departure flow is more perfect that that of return flow.

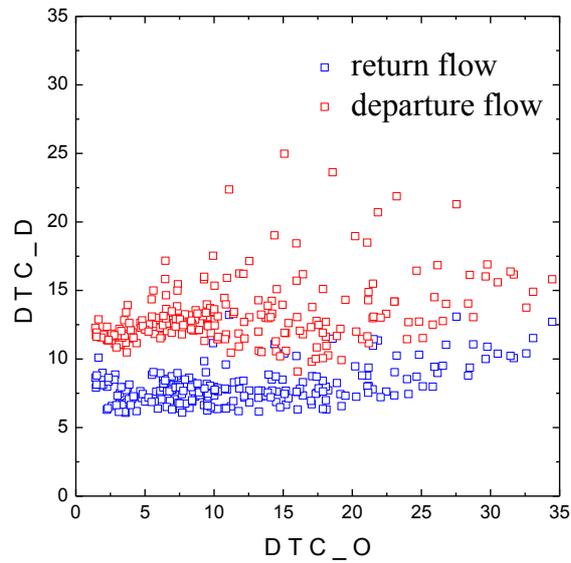

Figure 6. The average distance to center of the outflow of each station. Red:. the return flow, blue: the departure flow, The symbols are same with figure 4.

Figure 6 together with figure S1 in Supplementary Materials part 3 shows an interesting characteristc for round trips. In the averaging view, their intraday mobility is switching between the two rings. In the morning they depart from the periphery ring trap and travel to the inner ring trap, in the evening they move reversely.

All these facts imply one thing, the travel pattern of metro passengers is ring aggregation. Ring aggregation pattern is the second proof we find that the city structure is related to the human travel pattern.

## 4. Discussions

The ring aggregation pattern seems to be an exclusive characteristic of the metro transportation. We have examined the bicycle trip data of London and the taxi trip data of Shanghai, no ring aggregation and linear law can be found. The travel pattern for these two transportation modes is completely different from ring aggregation, as figure 7 shows, in each trip the destination and the origin are quite near each other, so we call this travel pattern "small regional pattern". And from figure 7 we see that the points of origin and destination indeed locate near the line $y = x$. In fact in the bicycle trip data of London we found a lot of trips with the same origin and destination, quite particular but clearly implying that the scale of movement is very small.

Why under these two transportation modes individuals travel to a place near where they depart? We think this may due that they are only suitable for short-distance travel, when the travel distance increases the travel cost (money, time, vigor, etc.) will rise significantly, then the destination is limited near the departure place. [27] also pointed out this remark. Different with bicycle and taxi, benefit from the subsidy from government, the metro system is much cheaper and is fairly suitable for middle-distance and long-distance travel, almost without cost limit. Especially for Beijing after paying 2RMB (about 0.3 dollars) a passenger can travel to any station, though as we know the longest path in this system contains 67 stations.

To the transportation mode which is suitable for even longer distance travel, such as flight and rail between cities, is the ring aggregation pattern also existing? For the lack of open and available data, we have not examined yet.

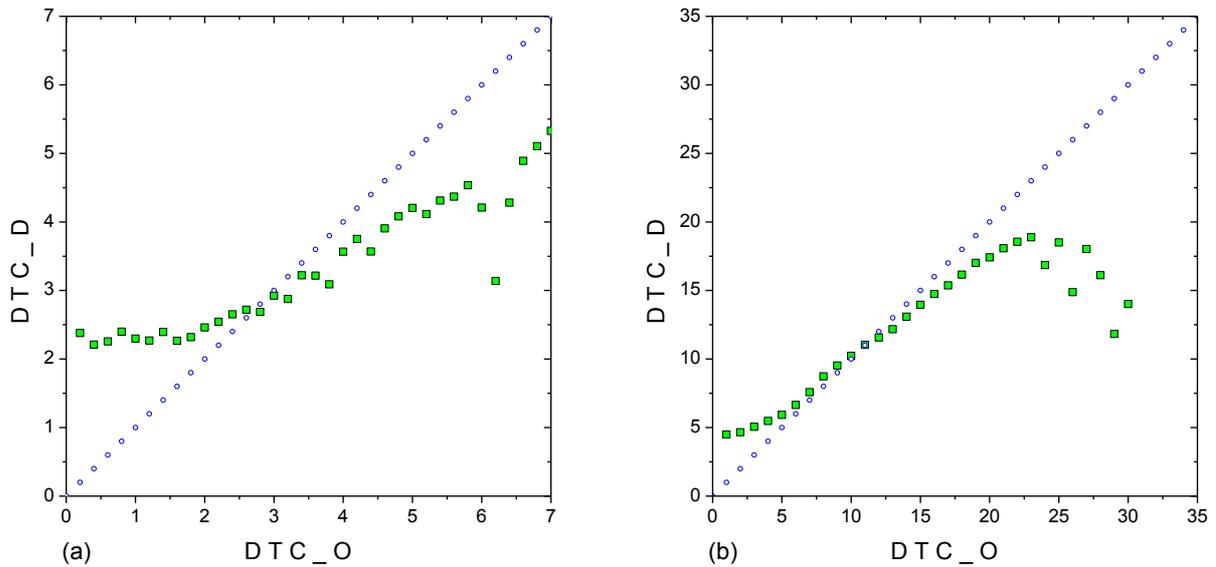

Figure 7. The average distance to center of the destination (DTC_D) with the distance to center (DTC_O). (a) London bicycle trip (b) Shanghai taxi trip. The symbols are same with figure 3

## 5. Summary and outlook

Though the society of city is a quite complex system, the rule determines human moving in city may be quite simple. In this work we find such a rule that a linear law exists between the individual's travel distance with his original distance to city center. The mechanism underlying is a travel pattern we called "ring aggregation", i.e. the daily movement of city passengers is just aggregating to a ring with the city center as its center. Interestingly, for the round trips the daily travel pattern can be regarded as a switching between the home ring at outer area with the office ring at the inner area. On one side this means city structure has remarkable relation with human mobility. To better understand human mobility it need to pay more considerations to city structure. On the other hand this line law is a new feature for the prediction of human mobility, which may improve the accuracy of prediction and be valuable for the widely used LBS applications.

# Acknowledgments

This work is partially supported by the National Natural Science Foundation of China under grant number 11302022, and the Fundamental Research Funds for the Central Universities under grant number 2013JBM048.